\pdfoutput=1
\documentclass[%
 reprint,
nofootinbib,
 amsmath,amssymb,
 aps, prd,
floatfix,
showkeys,
]{revtex4-1}

\usepackage{graphicx}
\usepackage{dcolumn}
\usepackage{bm}
\usepackage{amsmath}
\usepackage{slashed}

\begin{document}

\preprint{}

\title{A quintessence dynamical dark energy model from ratio gravity}

\author{Jackie C.H. Liu}
 \email{chjliu@connect.ust.hk}%

\affiliation{%
Institute for Advanced Study, Hong Kong University of Science and Technology, Hong Kong
}%
\affiliation{%
Department of Physics, Hong Kong University of Science and Technology, Hong Kong
}%

\date{\today}

\begin{abstract}
Based on the work of ratio gravity developed in 2018, which postulates the deformation of the cross ratio to associate with the physical model of gravity, we develop a mechanism to generate dynamical dark energy - a quintessence field coupled with gravity.  Such model causes the dark energy behaving differently in early and late time universe.  In the radiation-dominated-era and matter-dominated-era, the related analytical solutions of the quintessence field have an interesting property - starting as a constant field, then oscillating as the universe expands.  By Markov Chain Monte Carlo search of the parameter space with the local measurement (Type Ia supernovae) in the Bayesian framework, the probed range of $H_0$ (within $1\sigma$) overlaps the $H_0$ value inferred from Planck CMB dataset by $\Lambda$CDM model.

\end{abstract}

\maketitle

\section{\label{sec:level1}INTRODUCTION}

The accelerated expansion of the universe was discovered in 1998 by the observation of supernova \cite{Riess1998}. The puzzle of dark energy is one of the greatest problems in cosmology. Theorists propose different explanations; for example, a simple one is the cosmological constant, $\Lambda$, of the $\Lambda$CDM model, which is a widely accepted model because it is part of the theoretical framework of general relativity, GR, and consistent with the observations in great detail \cite{planck2015,planck2018}.  Alternative theories such as scalar field model of dark energy \cite{PRD084054} are also compelling cosmological models.

The measurements of the early universe and the local measurements (i.e. late-time universe) \cite{planck2018,Riess2016} probe the Hubble constant for different values with increasing accuracy so both results seem contradict, which is called the Hubble tension. It leads to many active research because of the potential implication for new Physics for our understanding of the gravity and cosmology \cite{PRD123525}.

There are several possible explanations to the tension such as statistical fluke, or the emerging spatial curvature effect from the cosmological model of the relativistic and nonlinear effect \cite{Bolejko2018}.  The study of the Planck data \cite{PRD063501} shows that, one of possible solutions to the Hubble tension, is that the equation of state of parameter of the dark energy, w, is not equal to -1.  
Another proposed resolution is the early dark energy model, EDE.  EDE models drive the expansion of the early universe (usually in radiation-dominated-era and/or matter-dominated-era), so that the expansions of the universe in the early era and late time era behave differently. In the work of Refs. \cite{axion,EDE}, one of the proposed axion models initially freezes at constant field value, then evolves to oscillate after the critical redshift so the EDE effect drives the early and late time of the expansion periods differently.  Poulin et al. \cite{EDE} analyse the models against the Planck CMB dataset and Type Ia supernovae dataset to show that the Hubble constant, $H_0$, probed by the model is consistent with the $H_0$ inferred by Planck CMB dataset. Furthermore, another proposed dark energy model is the acoustic dark energy model \cite{acousticDE}. It is related to the dark fluid that a scalar field converts its potential energy to kinetic energy during the matter-radiation equality.

The theory of ratio gravity, RG theory, is a newly developed theory \cite{Liu2018} that postulates the deformation of the cross ratio to associate with the physical model of gravity in the framework of Newman-Penrose formalism \cite{NP1962}. In the present work, we deploy a different approach from Ref.\cite{Liu2018} that we derive the physical models of fermion and scalar fields from the core equations of the RG theory in sections 2 and 3. We develop a generic framework to obtain the scalar field, which has the property of the symmetry breaking for vacuum expected value that is similar to the ordinary $\phi^4$ theory (in section 3).  

In this paper, we explore the interesting property of a quintessence dynamical dark energy model originated by the work of ratio gravity.  We study the prediction of the model in the the second part of the paper accordingly.

In section 4, by considering the scalar field as the quintessence dark energy model with CDM, we show the correspondence to the $\Lambda$CDM model in radiation-dominated and matter-dominated eras.  In such eras, we found the related analytical solutions of the field.  The solutions have an interesting property - starting as a constant field, then oscillating as the universe expands.  The models of axion dark energy \cite{axion,EDE} suggest similar scenario: the axion starts with "frozen" phase then transits to the oscillating phase.

In the last section, we perform the Markov Chain Monte Carlo search for the parameters of the qCDM model - quintessence and CDM model - with the Pantheon dataset - 1048 Type Ia supernovae (SNe Ia) \cite{pantheon}.  The probed Hubble constant is approximately equals to 67$\pm$4 km/s/Mpc.  Although we probe the model parameters by the dataset of late time universe, the probed range of $H_0$ (within $1\sigma$) surprisingly overlaps the $H_0$ deduced from $\Lambda$CDM by Planck CMB dataset \cite{planck2018}.  Due to the limited data analysis in this work, we make no conclusion to the possibility for resolving Hubble tension by this model.  Additional data analysis with more dataset such as BAO is recommended in the future work.

\section{Introduction to the Framework}

In this section, we introduce the principle of ratio gravity from the previous work \cite{Liu2018} and two core equations used throughout this paper. We develop a new framework that relies on the basic principle of Ref. \cite{Liu2018}, while modify the interpretation of the connection to gravity. We explain the difference at the end of this section.

The definition of the cross ratio over Riemann sphere is:
\begin{equation}
(z_1, z_2; z_3, z) = \frac{(z_3-z_1)(z-z_2)}{(z_3-z_2)(z-z_1)},
\nonumber
\end{equation}
where $z_1,z_2,z_3$ are complex numbers of the poles, and $z$ is the reference point over the Riemann sphere. A cross ratio consists of many equivalent representations while represents the same value. The arbitrariness of the same cross ratio allows four degrees of freedom because only four free parameters for three movable poles. By Ref. \cite{Yoshida}, one of the representations of the cross ratio is the hypergeometric differential equation of three regular singular poles.

Since the hypergeometric differential equation can be written as a second order linear differential equation in two-by-two matrix form and one can express it as the integrable system \cite{Cassidy}:
\begin{equation}
\partial_\mu Y =B_\mu Y.
\nonumber
\end{equation}
We can further introduce the gauge transformation to $Y$ and B matrices with re-definitions of B matrices to yield:
\begin{equation}
D_\mu Y=iB_\mu Y,							\label{eq:Y}
\end{equation}
where $D_\mu Y=\partial_\mu Y+i [\Lambda_\mu, Y]$ and $\Lambda_\mu$ are trace-less two-by-two Hermitian matrices.  Eq. (\ref{eq:Y}) is called $Y$ Equation. The D operator obeys Leibniz rule for derivation. One can further transform $Y$ equation by tensoring a Hermitian map to yield the form:
\begin{equation}
D_{ab} Y=iB_{ab} Y,							\label{eq:Y2}
\end{equation}
Eq. (\ref{eq:Y2}) is the original form of $Y$ equation in Ref. \cite{Liu2018} with spinor index $ab$.  Note that the Hermitian map can be associated with the metric components according to Newman Penrose formalism (NP formalism \cite{NP1962}).  The D operator can be defined in more general way as the form $D_\nu=f^{\nu\mu} \partial_\mu$\footnote{$f^{\nu\mu}$ are complex functions.}  as long as it is one-to-one corresponding to the D operator of $Y$ equation (\ref{eq:Y}) by associated automorphism.  In this paper, we use capital index, e.g. $A$, to denote the index abstractly to reserve the generalization for D operators.  

The gauge transformation of Eq. (\ref{eq:Y}) allows generating different representations of the $Y$ equation trivially so it is an automorphism - the transformed $Y$ equation and the original $Y$ equation are in the same space (i.e. the same mathematical structure).  Such automorphism allows we describe the same cross ratio with different equivalent representations.

Galois transformation is introduced to provide another form of transformation that respects the automorphism \cite{Liu2018}. The Galois transformation is defined by a Galois operator, $\hat{\rho}$, that obeys:
\begin{equation}
D \hat{\rho} = \hat{\rho} D,							\label{eq:gal}
\end{equation}
Eq. (\ref{eq:gal}) is called the \textit{Galois equation}. The definition of the Galois transformation and Galois equation originate from the Galois differential theory\footnote{Galois map $\pi :K\to K$, where $K$ is called differential field extension.} \cite{Cassidy}  - the theory studies the Galois groups of the differential equations. In the context of ratio gravity, we focus on how Galois equation provides the transformation of $Y$ equation and the related automorphism, so it requires no intensive knowledge of Galois differential theory.

In section 3, we derive the equation of motion from $Y$ equation and introduce a generic Galois operator that solves Galois equation and leads to the related scalar field equation.

Unlike the previous work \cite{Liu2018}, we use original framework of General Relativity instead of NP formalism. In previous work, the NP formalism is connected to set of Galois equations via the introduction of Bianchi constraints.

In this new framework (section 3), we first find the equation of motion of $Y$ equation (\ref{eq:Y}), and the related scalar field equation(s) from Galois equation (\ref{eq:gal}) to define the associated Lagrangian of matter - $\mathcal{L}_m$. Then, as the ordinary treatment of general relativity, we consider $\mathcal{L}_m$ as the source of gravity to define gravitational energy momentum tensor \cite{Wald1984}, i.e. using Einstein equation as the constraint equation to fix the degree of freedom of the metric.

In the context of RG, we use both Galois transformation and the continuous transformation of the metric elements by general relativity (GR) to find the space of transformed $Y$ equations, i.e. associated cross ratio representations. Note that the automorphism in the context of RG is not the same as the one in GR context - the general covariant transformations defined as the automorphisms of fibre bundles; RG requires the automorphisms applying to the space of $Y$ equations, i.e. the cross ratio representations.

\section{$Y$-Fermions and The Vacuum}

In order to find the equation of motion of $Y$ equation (\ref{eq:Y}) and the associated Galois equation (\ref{eq:gal}) in this section, we apply the gauge transformation to Eq. (\ref{eq:Y}), make use of Dirac equation, and find the related Galois equations. In the middle and last parts of this section, we explain the interpretation of the equation of motion of $Y$ equation and related Galois equation in the context of quantum field theory.  The purpose of this section is to define the physical models in the RG context under the framework of Lagrangian.

By applying a gauge transformation to Eq. (\ref{eq:Y}), one of the four components of $Y$ matrix can be gauged out because there are 3 degrees of freedom of the gauge in SU2. Therefore, there are four possible cases to choose the zeros of the components of $Y$ matrix. We classify them as four categories:
CAT 1$\to$4 of $Y$ matrix as follow
\begin{equation}
\left(
\begin{array}{cc}
 * & 0 \\
 * & * \\
\end{array}
\right),\left(
\begin{array}{cc}
 * & * \\
 0 & * \\
\end{array}
\right),\left(
\begin{array}{cc}
 * & * \\
 * & 0 \\
\end{array}
\right),\left(
\begin{array}{cc}
 0 & * \\
 * & * \\
\end{array}
\right). \nonumber
\end{equation}

In order to define the equation of motion by the eigen solutions of Eq. (\ref{eq:Y}), we introduce the parameterization to B and $\Lambda$ matrices:
\begin{align}
B_{\mu }&=\overset{1}{\Phi }\, p_{1 \mu } \hat{e} + \overset{2}{\Phi }\, p_{2 \mu } \hat{f} + \overset{3}{\Phi }\, p_{3 \mu } \hat{h} + \overset{4}{\Phi }\, p_{4 \mu } 1_2, \\
\Lambda _{a \mu }&=\overset{5}{\Phi }\, p_{5 \mu } \hat{e} + \overset{6}{\Phi }\, p_{6 \mu } \hat{f} + \overset{7}{\Phi }\, p_{7 \mu } \hat{h} ,
\end{align}
where we use matrix structure of $sl_2$ algebras ($\hat{e}, \hat{f}, \hat{h}$), $p_{a \mu }$ are dimensionless parameters, and $\overset{a}{\Phi }$ are complex functions.   We define $y$ for the column matrix of three dimensions to represent the non-zero components of $Y$ matrix.  We can re-write the $Y$ equation for CAT 1 $\to$ CAT 4 as the following form:
\begin{equation}
i \partial_\mu y= P_\mu y,							\label{eq:y}
\end{equation}
where $P_\mu$ are the three-by-three matrices.  There are constraints of \{$\overset{a}{\Phi}$\} needed to be satisfied to obtain Eq. (\ref{eq:y}); for instance, the explicit form of $P_\mu$ matrices for CAT 1 is 
\begin{equation}  \small
P_\mu = 
\left(
\begin{array}{ccc}
 -\overset{4}{\Phi } p_{4 \mu } & 0 & -\overset{3}{\Phi } p_{3 \mu } \\
 -\overset{2}{\Phi } p_{2 \mu } & \overset{3}{\Phi } p_{3 \mu }-\overset{4}{\Phi } p_{4 \mu }-2 \overset{7}{\Phi } p_{7 \mu } & -\overset{2}{\Phi } p_{2 \mu } \\
 -\overset{3}{\Phi } p_{3 \mu } & 0 & -\overset{4}{\Phi } p_{4 \mu } \\
\end{array}
\right),
\end{equation}
with $\overset{1}{\Phi}=\overset{5}{\Phi}=\overset{6}{\Phi}=0$.

The eigen-values of $P_\mu$ matrices of Eq. (\ref{eq:y}) for CAT 1,2,3,4 are \{$\pm \phi_1 p_\mu,\phi_2 p_\mu$\}, \{$\mp\phi_1 p_\mu,-\phi_2 p_\mu$\}, \{$\pm\phi_2 p_\mu,\phi_1 p_\mu$\}, and \{$\mp\phi_2 p_\mu,-\phi_1 p_\mu$\} respectively, where $p_\mu$ is the dimensionless momentum constructed by parameters $p_{a \mu }$, and $\phi_1$ and $\phi_2$ are the linear combinations of \{$\overset{a}{\Phi }$\}. Because of the first order differential operator of Eq. (\ref{eq:y}), we make use of Dirac equation in momentum space \{$L(p),R(p)$\} to obtain the equation of motion of $y$:
\footnote{The Dirac-slash-notation operators $\not\!p$ and $\not\!\partial$ contain the projector matrix for $l \leftrightarrow r$ implicitly.  $\not\!p  \phi^* L= m_y \phi R$ and $\not\!p \phi_2 R= m_y \phi^\dagger L$ are the constraints satisfied at real classical expected value of $\phi$.}
\begin{align}
&i \not\!\partial L= \not\!p \, \phi^* L= m_y \, \phi R, \nonumber\\
&i \not\!\partial R= \not\!p \, \phi_2 R= m_y \, \phi^\dagger L,	\label{eq:Dirac}			
\end{align}
where $m_y$ is the mass coupling, $L=L(p)l, R=R(p)r$ and $l^*$ are the doublet of first and second eigen-vectors, and $r$ is the third eigen-vector of $y$ respectively\footnote{The definition of $l^*,r$ to correspond to eigen-vectors of $y$ is only conventional because of the standard model framework; one can employ $l,r^*$ with the re-definition of $\phi \to \phi^*$.}, and we denote $\phi$ as the doublet form of \{$\phi_1,\phi_2$\}.  We naturally define the equation of motion of $y$ as the equation of motion for \textit{$Y$ fermion}. The equation of motion cannot be solved because the value of $\phi$ are not constrained, so we rely on Galois equation Eq. (\ref{eq:gal}) to fix it next.

In the context of quantum field, we interpret that the excitation of multiple $Y$ fermions by the Dirac Lagrangian associated with Eq. (\ref{eq:Dirac}) is corresponding to the set of multiple representations of the related $Y$ equation. It is merely the interpretation to relate the context of quantum field from the RG theory's perspective.

Cassidy \cite{Cassidy} defines the Galois map (automorphism $\pi$) to transform as $\pi: x \to y$, where $x$ and $y$ are elements of the space constructed by $Y$ matrix and the derivatives of $Y$ matrix, and $\pi \partial = \partial \pi$ as Eq. (\ref{eq:gal}). We define a generic Galois operator, $\hat{\rho}$, similarly:
$\hat{\rho}:= X^A D_A$,
where $\hat{\rho}$ satisfies Galois equation, i.e.
\begin{equation}
D_C \, \hat{\rho} \, Y = \hat{\rho} \,D_C \,Y, \label{eq:rho}	
\end{equation}
and $X^A$ are two-by-two-matrix functions because the operator D acts on two-by-two matrices. Certainly, one can define a more complicated Galois operator (e.g. higher derivative operator) with the potential cost of less solvability of Galois equation. To ensure $\hat{\rho}$ being associated with automorphism, we use the exponential map: $exp(\epsilon \hat{\rho})$, such that $Y$ equation (\ref{eq:Y}) transforms invariantly and infinitesimally by $\epsilon$ if Eq. (\ref{eq:rho}) is satisfied.

Because of the parameterization for $Y$ equation, the Galois equation (\ref{eq:rho}) can be generally expressed as (for CAT 1 and 4):
\small
\begin{align} 
&\omega^A \left(h_{\mathcal{C}} \partial _A\left(\overset{3}{\Phi }\right)-h_A \partial _{\mathcal{C}}\left(\overset{3}{\Phi }\right)\right)-\overset{3}{\Phi } h_A \partial _{\mathcal{C}}\left(\omega^A\right)=0,\nonumber\\
&\omega^A f_{\mathcal{C}} \partial _A\left(\overset{2}{\Phi }\right)-f_A \omega^A \partial _{\mathcal{C}}\left(\overset{2}{\Phi }\right)+\nonumber\\
&\overset{2}{\Phi } \left(2 \overset{7}{\Phi } \omega^A \left(f_A j_{\mathcal{C}}-j_A f_{\mathcal{C}}\right)-f_A \partial _{\mathcal{C}}\left(\omega^A\right)+2 \overset{3}{\Phi } \omega^A \left(h_A f_{\mathcal{C}}-f_A h_{\mathcal{C}}\right)\right)+\nonumber\\
&\mathit{f}^A \left(2 \overset{3}{\Phi } \overset{7}{\Phi } h_A j_{\mathcal{C}}-h_A \partial _{\mathcal{C}}\left(\overset{3}{\Phi }\right)+h_{\mathcal{C}} \partial _A\left(\overset{3}{\Phi }\right)\right)-\overset{3}{\Phi } h_A \partial _{\mathcal{C}}\left(\mathit{f}^A\right)=0 \label{eq:vac0}	,
\end{align}
\normalsize
where $j_A, h_A, f_A$ denote the parameters $i p_{7A}, i p_{3A}, i p_{2A}$ respectively, $\overset{4}{\Phi }$ is zero, and $X(x)^A = \omega(x)^A 1_2 + \mathit{f}(x)^A \hat{f}$.  For CAT 2 and 3, equations (\ref{eq:vac0}) are the same form by the transformation: $\mathit{f}(x)^A \to \mathit{e}(x)^A$ and $f_A \to e_A$.  We notice the equation above can be realized in a symbolic form as:
\begin{equation}
\partial \Phi = (\Phi) + (\Phi \Phi),
\end{equation}
where ($\Phi$) and ($\Phi \Phi$) denote the terms with the coefficients for the powers of ($\overset{a}{\Phi }$) and ($\overset{a}{\Phi } \overset{b}{\Phi }$) respectively, the equation is likely in the form for scalar field(s) with non-zero vacuum expected value, vev. The rest of this section is to prove this observation and construct the associated symmetry-breaking Lagrangian.

By applying the rest frame condition onto Eq. (\ref{eq:vac0}) for the dimensionless momentum, $p=(p_O,0)$, of $y$ fermion, we obtain:
\begin{align}
 \phi _1 \partial _{\mathcal{C}}\left(\mathit{f}^O\right)+\mathit{f}^O \partial _{\mathcal{C}}\left(\phi _1\right)=0, \nonumber \\
\mathit{f}^O \left(\mu  \phi _1^2+i \partial _O\left(\phi _1\right)\right)+i \phi _1 \partial _O\left(\mathit{f}^O\right)-i \mathit{f}\cdot{\nabla}\left(\phi _1\right)=0
\label{eq:vac1}
\end{align}
where $O$ denotes the time-axis-index, $O\neq C$, $\mathit{f}\cdot\nabla$ denotes directional derivate $\mathit{f}^A \partial_A$, and $\mu:=p_O$ is the dimensionless constant of the theory.  Eq. (\ref{eq:vac1}) is only a specific solution of Eq. (\ref{eq:vac0}) when we consider the case of singlet $\phi_1$ solution. (The doublet equation is not covered in this paper.)  In the rest of this paper, we denote $\phi$ as the singlet field.

In order to apply to a specific coordinate system for cosmology, we consider the $D_O$ operator by $D_O=a(t) \partial_t$ for FRW cosmology. The Laplacian of the time-dependent-only $\phi$ is\footnote{We use the degree of freedom of $\mathit{f}^C$, $C\neq O$, to remove the directional derivate term.}
\begin{equation}
\ddot{\phi }=-\frac{i \mu  \phi ^2 \dot{a}}{a^2}-\frac{2 \mu ^2 \phi ^3}{a^2}-\frac{3 i \mu  \phi ^2 \dot{\theta }}{a \theta }+\frac{2 \phi  \dot{\theta }^2}{\theta ^2}-\frac{\phi  \ddot{\theta }}{\theta },
\end{equation}
where $\theta$ denotes $\mathit{f}^O$,
and the associated Lagrangian is
\small
\begin{align} 
&-\frac{1}{2} g^{\mu \nu }\partial _{\mu } \chi _1  \partial _{\nu } \chi _1  -\frac{1}{2} g^{\mu \nu }\partial _{\mu } \chi _2  \partial _{\nu } \chi _2 - \nonumber \\
&(i \chi _1^3 -3 \chi _2 \chi _1^2 -3 i \chi _2^2 \chi _1 +\chi _2^3 )C + \nonumber\\
&\chi _1^2 A-\chi _2^2 A+\chi _1 \chi _2 B-\frac{\mu ^2 \left(\chi _1^4+\chi _2^4\right)}{2 a^2}+ \nonumber\\
&\chi _1^2 \left(\frac{3 \mu ^2 \chi _2^2}{a^2}+\frac{\mu  \chi _2 \dot{a}}{a^2}\right)-\frac{\mu  \chi _2^3 \dot{a}}{3 a^2},
\end{align}
\normalsize
where $\phi$ is expressed in real and imaginary parts: $\phi=\chi_1- i\, \chi_2$, $B=-2i A$, $C=\frac{\mu \dot{\theta}}{2a \theta}$ and $A=\frac{\dot{\theta}^2}{2 \theta ^2}-\frac{\ddot{\theta}}{4 \theta}$.
The Lagrangian above is not yet the physical model we look for.  The problematic complexness does not respect the Hermiticity of Lagrangian. So, we add the Hermitian conjudge terms. Because the Hermitian map of D operator (\ref{eq:Dirac}), the associated $Y$ fermion respects the symmetry of positive and negative energies. Therefore, we model the Lagrangian terms associated with the Galois equation in the current theory as $\mathcal{L}_\phi= \mathcal{L}_{+\phi} + \mathcal{L}_{-\phi}$ such that it respects positive-negative-vev-symmetry, just like the ordinary $\phi^4$ theory respects $Z_2$ symmetry. Obviously, this artificial symmetry breaks down if the $Y$-fermionic sector does not obey such symmetry. We consider this possibility to be the future development. Finally, we obtain the effective potential $V_\chi$:
\begin{equation}
\frac{\mu ^2 \left(\left(\chi ^*\right)^4+\chi ^4\right)}{4 a^2}-\frac{1}{2} \left(\chi ^2+\left(\chi ^*\right)^2\right) m_{\chi }(\theta)^2,
\label{eq:Vchi0}
\end{equation}
where $m_{\chi }(\theta)^2:=A$.

In this section, we show how to obtain the fermionic model of theory, i.e. $Y$ fermion, associated with $Y$ equation (\ref{eq:Y}) for CAT 1$\to$4, and the symmetry-breaking scalar field potential (\ref{eq:Vchi0}) by the generic Galois operator and Galois equation.

\section{The Quintessence Field and Cosmological Model}

In this section, we apply the framework of previous section to the application of cosmology - simplify the $\chi$ potential (\ref{eq:Vchi0}) to construct the quintessence field model that coupled to gravity. We show that such qCDM model corresponds to well-accepted $\Lambda$CDM with a derivation mechanism. Brief comparison to several established models \cite{axion,EDE} is covered.

The vacuum expected value of $\chi$ potential (\ref{eq:Vchi0}) is $\frac{m_{\chi }(\theta)}{\mu / a }$. It is not fixed because of the degree of freedom by $m_{\chi }(\theta)$, and it is dynamical as the scale factor varies. By requiring the vev of $\chi$ fixed, the $m_{\chi }(\theta)$ term should be proportional to $1/a$ so the vev becomes $\frac{m_{\chi }}{\mu }$, and the $\chi$ potential becomes
\begin{equation}
\frac{\mu ^2 \left(\left(\chi ^*\right)^4+\chi ^4\right)}{4 a^2}-\frac{1}{2 a^2} \left(\chi ^2+\left(\chi ^*\right)^2\right) m_{\chi }^2,
\label{eq:Vchi}
\end{equation}
where $m_{\chi }$ is a constant parameter.
In the rest of the paper, we consider the simple case to define the quintessence field, that $\chi$ is a real scalar field which recovers as $\phi^4$ potential at $a=1$, and it has the minimum degree of freedom needed to solve Friedmann equations.

The quintessence field potential is re-written as
\begin{equation}
\frac{\mu ^2 \chi ^4}{4 a^2}-\frac{m_{\chi }^2 \chi ^2 }{2 a^2}.
\end{equation}
We consider the Lagrangian of the quintessence\footnote{$Z_\chi$ is the overall coupling strength between gravity and $\chi$ field.}, $\mathcal{L}_q:=Z_\chi \mathcal{L}_\chi$, as the source of gravity to define gravitational energy momentum tensor for the quintessence field, and apply the usual variation on $\mathcal{L}_q$ to get the related density and pressure of the quintessence field\footnote{Prime denotes the derivative with respect to $a^2$.} \cite{Wald1984}:
\small
\begin{align}
\rho_q &= \frac{\mu ^2 q ^4}{4 a^2}-\frac{m_{\chi }^4}{4 a^2 \mu ^2}+\frac{\mu  q ^3 m_{\chi }}{a^2}+\frac{q ^2 m_{\chi }^2}{a^2}+\frac{1}{2} Z_{\chi } \dot{q}^2, \nonumber \\
p_q&=\frac{\mu ^2 q ^4}{4 a^2}-2 \mu ^2 q ^3 q '-\frac{m_{\chi }^4}{4 a^2 \mu ^2}+\frac{\mu  q ^3 m_{\chi }}{a^2}+\frac{q ^2 m_{\chi }^2}{a^2}- \nonumber \\& 
6 \mu  q ^2 m_{\chi } q '-4 q  m_{\chi }^2 q '+\frac{1}{2} Z_{\chi } \dot{q}^2, \nonumber
\end{align}
\normalsize
where we expand the $\chi$ field around vev, $\chi = \frac{m_{\chi }}{\mu } + q$, so we can deploy weak field limit next.  We notice that $V_\chi(a^2)$ contributes because of the variation, and absorb $Z_\chi$ factor for terms, $Z_\chi \mu ^2 \to \mu ^2$ and $Z_\chi m_{\chi }^2 \to m_{\chi }^2$. With the re-definition of the constants, the $Z_\chi$ factor is merely the rescaling factor for the time/energy scale between the quintessence and $\chi$ fields. We further assume the validity of \textit{weak-field-limit}, i.e. quadratic-terms-dominated, to yield:
\small
\begin{align}
\rho_q &= -\frac{m_{\chi }^4}{4 a^2 \mu ^2}+\frac{q ^2 m_{\chi }^2}{a^2}+\frac{1}{2} Z_{\chi } \dot{q}^2, \nonumber \\
p_q&=-\frac{m_{\chi }^4}{4 a^2 \mu ^2}+\frac{q ^2 m_{\chi }^2}{a^2}-4 q  m_{\chi }^2 q '+\frac{1}{2} Z_{\chi } \dot{q}^2.
\end{align}
\normalsize

In order to study the dark energy behavior, we define the dark energy density parameter, $\xi:=\Omega_q$, so
\begin{equation}
\xi=-\frac{\alpha ^4}{12 a^2 \lambda  H_0^4}+\frac{\alpha ^2 q ^2}{3 a^2 H_0^4}+\frac{\lambda  Z_{\chi } \dot{q}^2}{6 \mu ^2 H_0^4}
\end{equation}
where $\mu ^2 := \frac{\lambda  M_p^2}{H_0^2}$ and $m_{\chi }:=\frac{\alpha  M_p}{H_0}$ for the ease of parameter probing next.  By Friedmann equations (without curvature k and cosmological constant terms), we have
\small
\begin{align}
&\dot{\xi}= \nonumber \\
& H (\frac{\alpha ^6}{\lambda }+\frac{2 a F H_0^2 \ddot{a}+F H_0^4 a^{2-r} (4 \xi  a^r+\Omega_{m} (3 w+1))}{\mu ^2 q}-4 \alpha ^4 q^2)/ \nonumber \\
&(6 a^2 \alpha ^2 H_0^4), \nonumber
\end{align}
\normalsize
where $F := (2 \alpha ^2 \mu ^2 q +a^2 \lambda  Z_{\chi } \ddot{q})$, $w$ is the equation of state of the matter component, and $r=3,4$ for matter and radiation-dominated eras respectively. We found if we define the equation of motion for q
\begin{equation}
\ddot{q} = -\frac{2 \alpha ^2 \mu ^2 q }{\mathit{a}^2 \lambda  Z_{\chi }} 	\label{eq:qEoM}
\end{equation}
then, when $F=0$, the weak field limit is valid up to a long period of cosmological time span.  The Friedmann equations together with Eq. (\ref{eq:qEoM}) can be expressed as
\begin{equation}
\xi =\frac{H^2}{H_0^2}- \frac{\Omega _m}{a^{r}},
\dot{\xi}=\frac{\alpha ^2 H \left(\alpha ^2-4 \lambda  q^2\right)}{6 a^2 \lambda  H_0^4},
\ddot{q}=-\frac{2 \alpha ^2 \mu ^2 q}{a^2 \lambda  Z_{\chi }}. 		\label{eq:Cos}
\end{equation}
Eq. (\ref{eq:Cos}) are the equations of the qCDM model of the quintessence theory.

The $\Lambda$CDM model is effective and supported by many observations. We need to verify the validity of qCDM model analytically against $\Lambda$CDM model. It is clear if $\dot{\xi}$ is zero, i.e. $\alpha ^2=4 \lambda  q^2$, then we recover the case of the cosmological constant and the first equation of (\ref{eq:Cos}) is simply the Friedmann equation with the cosmological constant so the $\Lambda$CDM correspondence is satisfied. Therefore, $q$ must be approximately constant and equals to $\frac{\alpha }{2 \sqrt{\lambda }}$ in order to justify the validity of $\Lambda$CDM correspondence. It can be achieved by the quintessence field staying approximately constant for long period of time or oscillating very slowly.

In both radiation-dominated-era and matter-dominated-era, i.e. $Log(a) \propto Log(t)$, we can solve $q$ analytically.  In matter-dominated-era,
\begin{equation}
q= (c_2 + c_1 \omega T) \sin (\omega T)+(c_1-c_2 \omega T) \cos (\omega T),
\end{equation}
where $\omega = \frac{3 \sqrt{2} \alpha  \mu }{\sqrt{\lambda } \sqrt{Z_{\chi }}}$, $T=\sqrt[3]{t/t_0}$, $c_1, c_2$ are integration constants, and $t_0$ is the present time, $a(t_0)=1$; in radiation-dominated-era,
\begin{equation}
q= c_3 T \, J_1(\omega T)+c_4 T \, Y_1(\omega T), \label{eq:qRad}
\end{equation}
where $J_1$ is a Bessel function of the first kind, $Y_1$ is a Bessel function of the second kind, $\omega = \frac{2 \sqrt{2} \alpha  \mu }{\sqrt{\lambda } \sqrt{Z_{\chi }}}$, $T=\sqrt{t/t_0}$, and $c_3, c_4$ are integration constants.  Both analytical expressions lead to the constant mode as $t\to0$, so in the early universe, the quintessence field is asymptotically constant and later evolves to oscillate.

The Eq. (\ref{eq:qRad}) is similar to the scalar field model in Ref. \cite{PRD103528} that showed ultra-light scalar fields affect growth of structure in the Universe as well as the expansion rate.  In the limit that, $a(t) \propto t^{p}$, i.e. in both radiation-dominated and matter-dominated eras, the analytical form of the axion-like particles is similar to (but not the same as) Eq. (\ref{eq:qRad}).

\begin{figure}[ht]
\includegraphics[width=250pt]{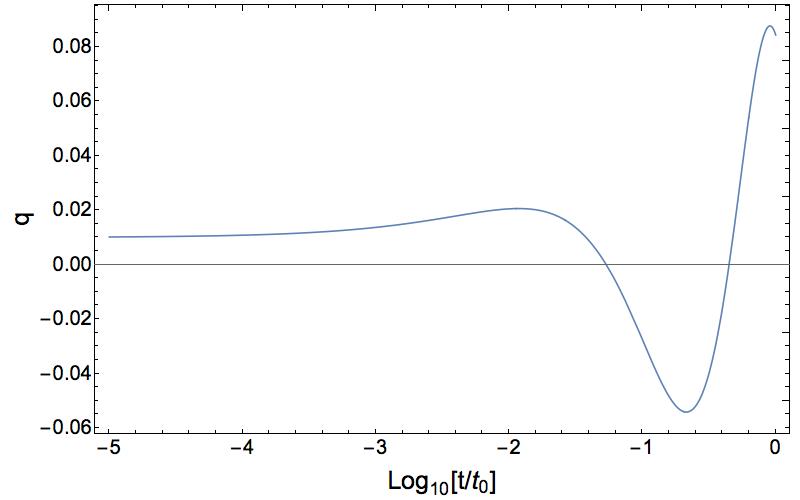}
\caption{\label{fig:qMat} 
Sample plot of the analytic formula of quintessence in the matter-dominated-era: $\alpha= 2,\lambda=1,Z_{\chi} =1$.  $t_{0}=1$ is the present time.}
\end{figure}

Given that the quintessence is nearly constant, $q_c$ , $\xi$ is solved
\begin{equation}
\xi = \xi_0-\frac{C}{a^2},   \label{eq:Ck}
\end{equation}
where $\xi_0$ is an effective cosmological constant term and $C = \frac{\alpha ^4-4 \alpha ^2 \lambda q _c^2}{12 H_0^4 \lambda }$ as a constant parameter associated with the term scaling as the spatial curvature (i.e. $a^{-2}$).  We have shown by the analytical form of the quintessence field that the near-constant-approximation is applicable in the early universe (in radiation-dominated-era and matter-dominated-era), and then the field oscillates shown in FIG.~\ref{fig:qMat}; the dark energy density parameter $\xi$ has an effective cosmological constant term in Eq. (\ref{eq:Ck}).

Interestingly, the models of dark energy \cite{axion,EDE} suggest similar scenario (but not exactly the same): the axion starts with "frozen" phase as the cosmological constant then transits to the oscillating phase, FIG.1 of Ref. \cite{axion}.

\section{Data Analysis}

In our data analysis section, we first identify a smaller set of parameter spaces of Eq. (\ref{eq:Cos}) that reduces the qCDM model to an effective and simplified qCDM version; following the same procedure as in \cite{EDE,PRD123507}, we use the Pantheon dataset of 1048 SNe Ia \cite{pantheon} to probe the model parameters with $\Omega_m$ dominated by matter component only.

The original parameters of qCDM are \{$H_0, \Omega_m, Z_\chi, \lambda, \alpha, q_0, q_1$\}, where $q_0$ and $q_1$ denote the $q(t_0)$ and $\dot{q}(t_0)$ of the present day.  We fix $Z_\chi=\mu^2$ because it allows that $\lambda$ is in scale within unity and $\alpha$ is in scale of $H_0$.  We also assume the effectiveness of $\Lambda$CDM in the present day such that $\dot{\xi}(t_0)$ and $q_1$ are effectively zero so $4q_0^2=\alpha^2\lambda$.  The parameters of our simplified qCDM model are \{$H_0,\Omega_m, \lambda, \alpha$\}.\footnote{The base unit of $\alpha$ and q is set as 73.9 km/s/Mpc.}

\begin{figure}[t]
\includegraphics[width=250pt]{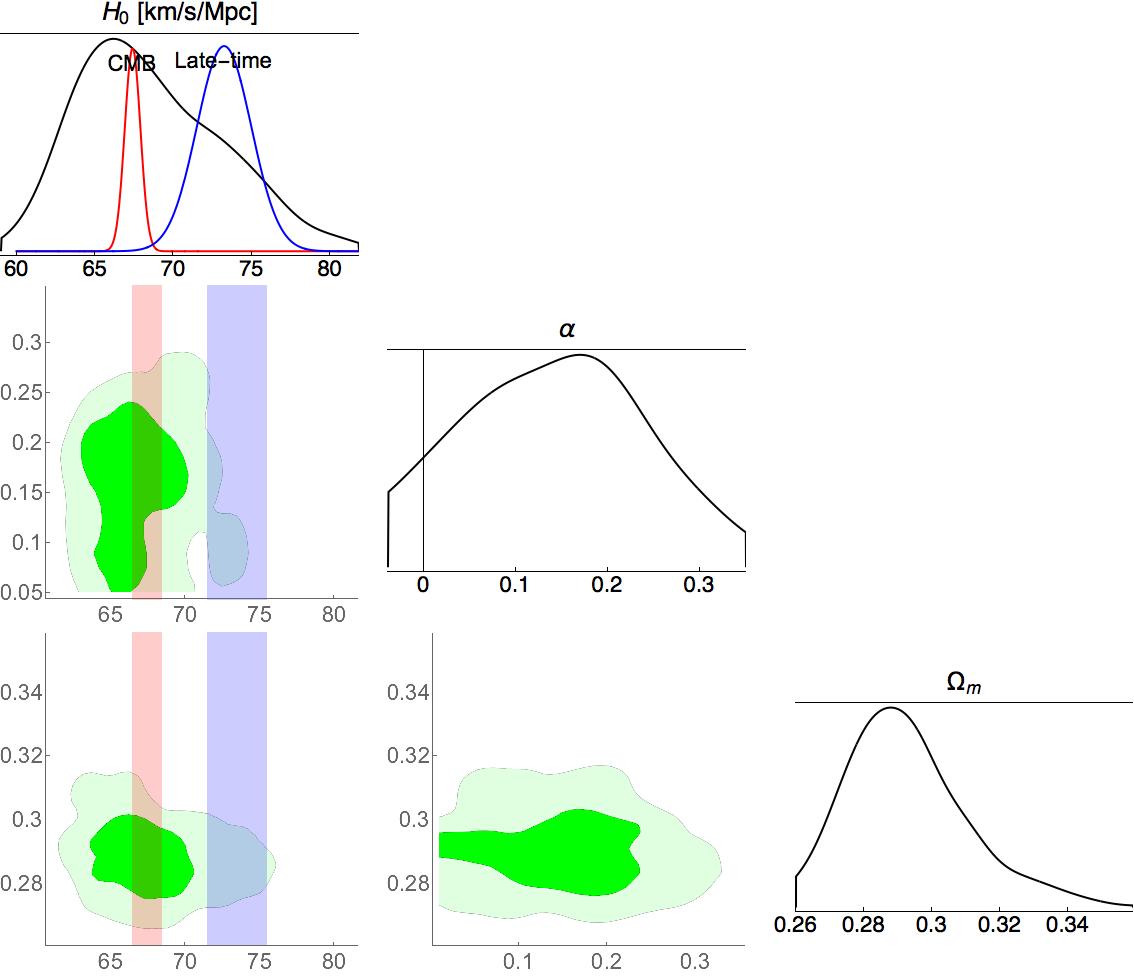}
\caption{\label{fig:posterior} 
1D and 2D posterior distributions of $H_0, \alpha, \Omega_m$ probed by the late time measurement (Pantheon dataset of 1048 SNe Ia).  The mean $H_0$ ($\pm1\sigma$) of qCDM model is approximately equals to 67$\pm$4 km/s/Mpc; the referenced $H_0$ of Planck \cite{planck2018} and the referenced $H_0$ of the late time measurements \cite{Riess2016} are shown in red and blue respectively.}
\end{figure}

The sampling by the numerical solving for Eqs. (\ref{eq:Cos}) consumes the computation resources seriously; therefore, we impose the prior-assumption.  By applying Markov Chain Monte Carlo, MCMC, parameters searching\footnote{An open sourced MCMC Mathematica package with modification for the our model (https://github.com/joshburkart/mathematica-mcmc).} and assuming the flat priors\footnote{We use the preferred prior from the known cosmological parameter: $0.24 < \Omega_m < 0.36$, $57 < H_0 < 85$ km/s/Mpc.} on \{$H_0,\Omega_m, \lambda, \alpha$\}, we perform the initial MCMC exploration on binned data (40 data-points), then we identify the preferred prior region for $\alpha \leq 0.25$, whereas the small-valued $\alpha$ is theoretically suggested because $\xi$ becomes effectively the cosmological constant as $\alpha \to 0$ by Eq. (\ref{eq:Cos}).  The full MCMC run on the Pantheon dataset (1048 SNe Ia) \cite{pantheon} shows the region of convergence from the 1D and 2D posterior distributions in FIG.~\ref{fig:posterior} with Gelman-Rubin criterion $R - 1 < 0.024$.

The $\chi^2_{min}$ in the MCMC run is $1035.7$, that is slightly better than $1036.5$ from the $\chi^2_{min}$ of $\Lambda$CDM probe \cite{PRD123507}.  The mean $H_0$ ($\pm1\sigma$) of qCDM model is 67$\pm$4 km/s/Mpc, that overlaps the estimated range (67.4$\pm$0.5 km/s/Mpc) by Planck measurement \cite{planck2018}.  We found the acceptable range of parameter $\lambda$ is board; we report the mean-best-fit-parameters\footnote{Taking the mean values of parameters of the top 2\% best-fit parameters within the 1$\sigma$ region of $H_0$, $\Omega_m$, and $\alpha$.} as \{$H_0=$67.2 km/s/Mpc$, \Omega_m=0.285, \lambda=0.698, \alpha=0.132$\}.

Finally, we use the Eq. (\ref{eq:Ck}) and the mean-best-fit-parameters to check briefly if the $C$ term of qCDM model is consistent with the CMB power spectrum.  Without a complete probe of qCDM parameters, we only change the $\Lambda$ term, cosmological constant of $\Lambda$CDM, by the Eq. (\ref{eq:Ck}).  The value of $q_c$ is $0.071$, and the order of magnitude of $C$ is $-4.9$.  We obtain the values of $\chi^2$ of CMB power spectrum against the dataset of \textit{PlanckTT} and \textit{WMAPTT}\footnote{\textit{PlanckTT}  dataset is corresponding to 111 data-points of $2 \leq l \leq 2000$, and \textit{WMAPTT} dataset is corresponding to 45 data-points of $2 \leq l \leq 1150$ from the open-source-CMB tool (http://www2.iap.fr/users/pitrou/cmbquick.htm)}.  The $\chi^2$ of PlanckTT and WMAPTT are only shifted by 0.02 and 0.07 respectively ($\chi^2$ for PlanckTT and WMAPTT are 493.37, 42.54 respectively).  However, the complete parameter probe of qCDM against CMB power spectrum is not covered in this work.

\section{Discussion}

In this work, we introduce the framework of ratio gravity that postulates the transformation of cross ratio is related to different representations of the associated fermion and scalar models.  The theory provides the mechanism to generate the symmetry-breaking scalar fields naturally, that leads to the quintessence field to drive the dark energy behaving dynamically.

The presented qCDM model can reproduce the $\Lambda$CDM model with a derivation mechanism.  The data analysis of the model with the supernovae dataset suggests $H_0$ = 67$\pm$4 km/s/Mpc, which is aligned with the latest Planck observation \cite{planck2018}.  Yet, a further analysis with complete set of qCDM parameters against CMB and BAO dataset is suggested.  

The theory can be extended to the domain of complex singlet and doublet models of the scalar field.  As the mass scale of the quintessence field is in $H_0$ suggested in the section 5, the possibility of light-massive boson because of the complex phase of the singlet model is worth to be studied.

\begin{acknowledgments}
JCHL would like to thank Professor Wang Yi and Professor Henry Tye for the valuable comments and advices.  This research was conducted using computational resources at the Institute for Advanced Study, Hong Kong University of Science and Technology.
\end{acknowledgments}


\bibliography{export}

\end{document}